\documentclass[doublecol]{epl2} 

\title{Nonlinear delocalization on disordered Stark ladder}
\shorttitle{Nonlinear delocalization on disordered Stark ladder} 
 
\author{Ignacio Garc\'ia-Mata\inst{1,2,3} \and Dima L.Shepelyansky\inst{1,2}}
\shortauthor{I.Garc\'ia-Mata, D.L.Shepelyansky}

\institute{                    
  \inst{1} Universi\'e de Toulouse, UPS, 
Laboratoire de Physique Th\'eorique (IRSAMC),  F-31062 Toulouse, France\\
  \inst{2} CNRS, LPT (IRSAMC), F-31062 Toulouse, France \\
  \inst{3} Departamento de F\'isica, Lab. TANDAR, Comisi\'on Nacional de Energ\'ia 
  At\'omica, Av. del Libertador 8250, \\
  1429 Buenos Aires, Argentina \\
\\
  \inst{} Dated: March 12, 2009
}

\pacs{05.45.-a}{Nonlinear dynamics and chaos}
\pacs{63.50.-x}{Vibrational states in disordered systems}
\pacs{03.75.Kk}{Dynamic properties of condensates; 
                collective and hydrodynamic
                excitations, superfluid flow }

\abstract{We study effects of weak nonlineary on
localization of waves in disordered Stark ladder
corresponding to propagation in presence of disorder and
a static field. Our numerical results show that
nonlinearity leads to delocalization with subdiffusive 
spreading along the ladder. 
The exponent of spreading remains close to its value
in absence of the static field.
}

\begin{document}

\maketitle

Anderson localization leads to suppression of diffusive propagation
of linear waves in systems with disorder \cite{and1958}.
In one and two dimensions all states are exponentially localized
\cite{and1979,lee}. For classical waves
nonlinearity is naturally present
and it is important to understand how it affects
localization in a random media \cite{kivshar}.
At first glance it seems that a spreading in space
leads to an effective decrease of nonlinearity
and hence persistence of localization
\cite{doucot}. On the other hand it was argued that
nonlinear resonances remain overlapped and
localization is destroyed by a moderate nonlinearity
which leads to a subdiffusive spreading in space
at asymptotically large times \cite{dls1993}.

Recent experimental progress with nonlinear photonic lattices
\cite{segev,lahini} and Bose-Einstein condensates (BECs) in optical lattices 
\cite{aspect,inguscio} with disorder  generated 
a renewal of significant theoretical interest to this problem
(see 
\cite{shapiro,pavloff,flach2008,dls2008,arkadyepl,aubryarxiv,flach2009,dls2009}
 and Refs. therein). Similar type of problems appear also 
for energy propagation
in disordered molecular chains 
\cite{flach2008,dhar} that enlarge a field of possible
applications. In addition the problem of interplay of 
localization and nonlinearity represents an interesting
mathematical problem of stability of pure point spectrum
with respect to nonlinear perturbations which 
led to recent mathematical studies \cite{wang,fishman}. 

The numerical studies are mainly done for 
the discrete Anderson nonlinear Schr\"odinger equation (DANSE)
showing that the wave packet width $\Delta n$ spreads at large times $t$
in a subdiffusive way with $(\Delta n)^2 \propto t^\alpha$
and an exponent $\alpha \approx 0.3 - 0.4 $ for system dimension $d=1$
\cite{dls1993,dls2008,flach2009} and $\alpha \approx 0.25$
for $d=2$ \cite{dls2009}. The theoretical estimates
give $\alpha=2/5$ \cite{dls1993} and $\alpha=1/4$ \cite{dls2009}
respectively. A noticeable difference between the theory estimates
and numerical value $\alpha \approx 0.3$  for $d=1$
is argued to be related with a specific properties
of 1d Anderson model \cite{dls2009} but further clarifications
of this point are required  (see e.g. \cite{flach2009}).

In this work we address a new type of question for the DANSE model:
how a static field force affects the properties of nonlinear delocalization?
Such a force is experienced by BECs in a gravitational field
or effectively in a magnetic field gradient. It can be also effectively
created by an acceleration of the optical lattice as a whole. 
This creates an effective Stark ladder which already has been realized
in experiments with cold atoms \cite{raizen}.
In absence of nonlinearity  a weak static field does not
significantly affect the localization while at strong fields
the localization length is significantly reduced
since less states are energetically available for hopping
over the ladder (see e.g. recent studies \cite{kolovsky}
and Refs. therein). It is not so obvious what are the effects
of nonlinearity in such a system since the nonlinear term is local
and is small compared to an energy variation
for large displacements along the ladder.

To answer the above question we study numerically the 
Stark DANSE model described by the equation
\begin{equation}
i \hbar{{\partial {\psi}_{n}} \over {\partial {t}}}
=(f n + E_{n}){\psi}_{n}
+{\beta}{\mid{\psi_{n}}\mid}^2 \psi_{n}
 +V ({\psi_{n+1}}+ {\psi_{n-1})}\; ,
\label{eq1}
\end{equation}
where $f$ is the strength of the Stark field.
At $f=0$ the model is reduced to the usual DANSE
studied recently in \cite{flach2008,dls2008,flach2009,aubryarxiv}. 
We fix the units as $V=\hbar=1$ and choose a typical set of parameters
used here as $W=4, \beta=1$. 
On-site energies $E_n$ are randomly and 
homogeneously distributed in the interval $-W/2 < E_n < W/2$.
Then the localization length in the middle of energy band
is $\ell \approx 6$ at $f=0$. The numerical integration was done
by the split operator scheme described in \cite{dls2009}.
Such a symplectic integration with
an integration time step $\Delta t =0.1$ and $0.01$
gives the energy conservation with accuracy $3\%$ and $1\%$
for $t \leq 10^7$ in a presence of strong field $f \leq 2$.
The total number of states was fixed at $N=256$, we used 
averaging over $N_d=15$ disorder realisations.  
The finite value of $\Delta t$ generates high frequency 
equidistant harmonics
with frequency spacing
$2\pi/\Delta t$. At $f=0$ these frequencies are located outside
of energy band while at $f>0$ they, in principle, 
may give resonant transitions. However, 
at small $\Delta t$ the distance between such resonant states
is much larger than the localization length $\ell$
and the matrix elements in such cases are exponentially small
and do not affect the behavior of the system
with variation of $\Delta t$ (see Fig.1, inset). 
The integration scheme conserves exactly the total probability.

\begin{figure}[htb]
\begin{center}
\includegraphics[width=8cm]{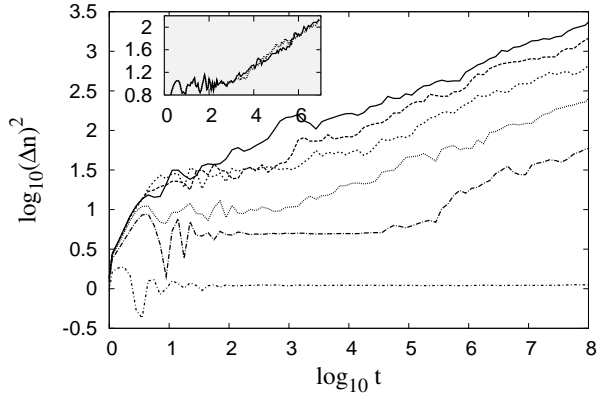} 
\caption{Dependence of the second moment 
$(\Delta n)^2$ of probability distribution on time $t$
for various values of static field $f$ and 
one fixed disorder realisation. Curves from top to bottom at 
$t=10^8$ are for $f=0., 0.1, 0.25, 0.5, 1., 2.$
and $\beta=1, W=4, N=256, \Delta t=0.1$. Inset shows data for $f=0.5$
obtained with integration steps $\Delta t=0.1$ (solid curve) and 
$\Delta t=0.01$ (dotted curve). To suppress  
fluctuations time average was made on logarithmic scale 
inside intervals $\Delta(\log_{10}t)=0.1$. Initial
state is one lattice site $n=0$ with energy in the middle of the band.
\label{fig1}
}
\end{center}
\end{figure}

\begin{figure}[htb]
\begin{center}
\includegraphics[width=8cm]{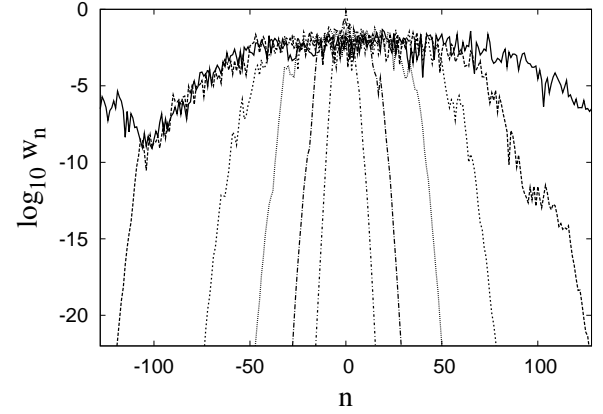} 
\caption{Probability distribution $w_n=|\psi_n|^2$ over ladder site $n$
at time $t=10^8$  for  $f=0., 0.1, 0.25, 0.5, 1., 2.$
(curves from small (inside) to large (outside) values of $|n|$).
Parameters are the same as in Fig.~\ref{fig1}.  
\label{fig2}
}
\end{center}
\end{figure}

A typical dependence of the second moment 
$(\Delta n)^2$ 
of probability distribution $w_n$
on time $t$ is shown in Fig.~\ref{fig1}
for various values of $f$.
The distribution of on-site probabilities
$w_n$ at a final time $t=10^8$ is shown in Fig.~\ref{fig2}.
The fits of  data of Fig.~\ref{fig1} show
an unlimited subdiffusive growth $(\Delta n)^2 \propto t^\alpha$
with the exponent $\alpha  =0.275 \pm  0.006$,
$0.291 \pm 0.005$, $0.276 \pm 0.006$, $0.272 \pm 0.004$,
$0.269 \pm 0.016$ for $f=0., 0.1, 0.25, 0.5. 1.$
respectively. The fits are done in the interval
$5 \leq \log_{10}t \leq 8$ for the last case and
$3 \leq \log_{10}t \leq 8$ for all other cases.
The data show no significant variation of $\alpha$
with $f$ even if one realization at finite time
may have noticeable fluctuations of $\alpha$ 
being larger than a formal statistical error 
(see below). At large values of $f=1., 2.$
the localization length $\ell$ becomes 
rather small (it can be estimated as 
$\ell \approx \sqrt{(\Delta n(t=1000))^2}$)
and during a long time interval there is not spreading over the ladder.
For $f=1.$ the growth appears at $t \geq 10^5$.
There is no visible growth for all computational times
for $f=2$. We interpret this as very low transition rates
over localized states in the case of small localization length 
$\ell \approx 1$ at $f=2$. It remains unclear if localization 
persists or disappears for such small $\ell$ at very large times.
For the cases with clear delocalization at $f=0., 0.1, 0.25, 0.5$
the distribution of $w_n$ over $n$ have a form of 
homogeneous ``chapeau'' centered near the initial state $n=0$;
its width  grows with time, approximately 
in agreement with the second moment growth
(see Fig.~\ref{fig2}). 

\begin{figure}[htb]
\begin{center}
\includegraphics[width=8cm]{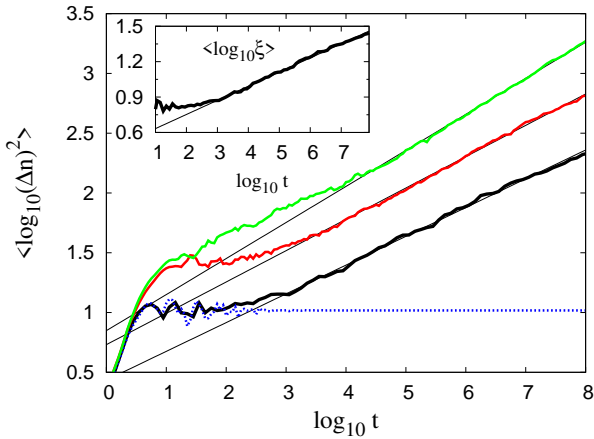} 
\caption{(Color online) Dependence of the second moment
$(\Delta n)^2$ on time $t$, average is done over 
$N_d=15$ disorder realisations.
The curves from top to bottom at $t=10^8$ are for 
$f=0., 0.25, 0.5$ at $\beta=1$ and
$f=0.5$ at $\beta=0$. 
The fits, shown by thin straight lines,
give the exponent of growth at $\beta=1$:
$\alpha=$ $0.302 \pm  0.001 (f=0)$ , $0.262 \pm 0.001 (f=0.25)$,
$0.241 \pm 0.002 (f=0.5)$.
Inset shows the dependence of participation ratio $\xi$
on time for $\beta=1, f=0.5$ (thick curve), the straight line
shows the fit dependence with the exponent 
$\nu=0.120 \pm  0.001$. Other parameters are as in Fig.~\ref{fig1}.
\label{fig3}
}
\end{center}
\end{figure}

\begin{figure}[htbp]
\begin{center}
\includegraphics[width=8cm]{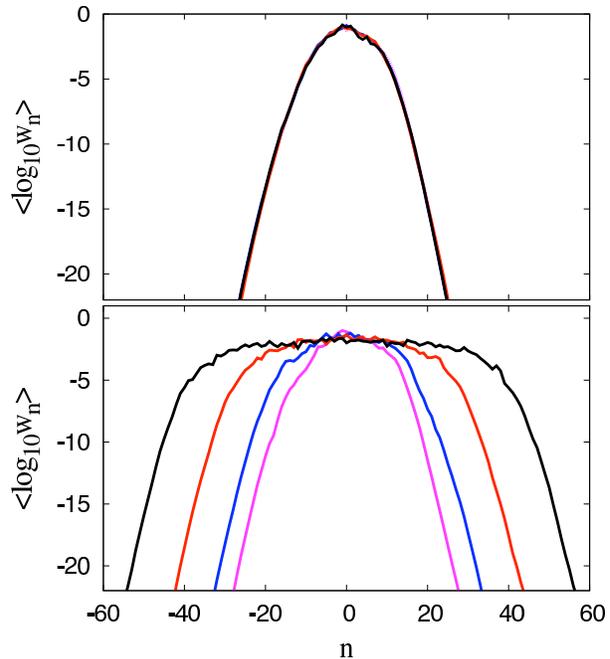}  
\caption{(Color online) Averaged probability distribution 
$w_n$ as a function of the lattice site $n$ for the cases of
Fig.~\ref{fig3} at $f=0.5$ (average is done over the same 15 disorder realisations). 
Top panel is for $\beta=0$ and bottom panel is for
$\beta=1$.
Curves are drown from inside (small $|n|$) to outside
(large $|n|$) for  
$t=10^2$ (magenta), $10^4$ (blue), $10^6$ (red) and 
$10^8$ (black). For $\beta=0$ (top panel) 
the state is almost the same for all times from $t=10^2$ to $t=10^8$. 
\label{fig4}
}
\end{center}
\end{figure}

To obtain more statistics we perform averaging over $N_d$
disorder realisations. The data are presented in Fig.~\ref{fig3}
for the second moment and in Fig.~\ref{fig4} for the probability distribution
for $\beta=0; 1$. At $f=0$ we obtain the exponent $\alpha =0.302$
which is comparable with the values $\alpha \approx 0.33$
found in previous studies \cite{dls2008,flach2009,dls2009}.
The value of $\alpha$ decreases by about $10\%$
when the static field is increased up to $f=0.5$.
This decrease is well visible even if it is not very large
and is comparable to the statistical variations of $\alpha$
at $f=0$ discussed above. We also computed
the dependence on time for the participation ratio
$\xi = 1/<\sum_n w_n^2>$. It can be characterized by the dependence
$\xi \propto t^\nu$ with the exponent $\nu=0.120 \pm 0.001 (f=0.5)$,
$0.131 \pm 0.001 (f=0.25)$, $0.159 \pm 0.002 (f=0)$.
These values are compatible with the usual relation
$\nu = \alpha/2$.

The averaged probability distributions for the cases of Fig.~\ref{fig3} 
at $f=0.5$ are shown in Fig.~\ref{fig4} at different moments of time.
For $\beta=0$ the distribution is localized being frozen in time.
It drops faster than exponential due to the presence of
static field showing a qualitative difference between
Stark localization and usual exponential Anderson localization.
For $\beta=1$ the probability spreads over the whole
lattice forming a homogeneous plateau in the center.
The interesting property of this distribution is its approximate
symmetry with respect to the initial state $n=0$.
It is clear that the conservation of energy imposes such a
symmetric spreading. Indeed, the width of the plateau is 
approximately $\delta n \approx 80$ and the energy variation
on such a distance is $\delta E \approx f \delta n \approx 40$
that is much larger than the energy band $ B \approx 6$
at $f=0$. Due to energy conservation at $|f|>0$ the spreading
can continue unlimitedly only in approximately symmetric way.
This excludes the possibility of a compact packet which moves
over a lattice on larger and larger distances in some 
stochastic way (as discussed in \cite{flach2008}).
A quasi-symmetric spreading seen in Fig.~\ref{fig4} looks rather 
natural in view of total energy conservation at $|f|>0$.
However, it raises an interesting problem of 
{\it statistical entanglement} of  probabilities $w_n$
on opposite ends of the plateau. Indeed, 
on such a distance the probabilities seems to be uncorrelated
since they are separated by many localization lengths $\ell$
of the linear problem. Nevertheless the propagation on these far ends
goes in a correlated way since the total energy
$E_{tot} \approx \sum_n f n w_n$ is exactly conserved. 

In \cite{dls1993,dls2008,dls2009}
it was argued that an infinite spreading is possible since the
nonlinear frequency shift $\delta \omega \sim \beta / \Delta n$
remains comparable with the 
frequency spacing $\Delta \omega \sim 1/\Delta n$
between frequencies of excited 
$\Delta n$ modes. On the Stark ladder this condition seems to be 
violated at large $\Delta n$ 
since $\Delta \omega \sim f \gg \delta \omega \sim \beta/\Delta n$.
However, the situation is more subtle. Indeed, we can write DANSE (1)
in the basis of linear eigenmodes.
The time evolution amplitudes $C_m$ in this basis is described by
equation (see e.g. \cite{dls1993,dls2009}):
\begin{equation}
i {{\partial C_{m}} \over {\partial {t}}}
=(f m + \epsilon_{m}) C_{m}
+ \beta \sum_{{m'}{m_1}{m_1'}}
V_{{m}{m'}{m_1}{m_1'}}
C_{m'}C^*_{m_1}C_{m_1'} 
\label{eq2}
\end{equation}
where $m$ marks the center of eigenmode inside the ladder
and eigenenergies $\epsilon_m$ are randomly distributed inside
the energy band width of approximately the same size $B \sim 4$
as at $f=0$. $V_{{m}{m'}{m_1}{m_1'}} \sim l^{-3/2}$ are the transition
matrix elements induced by nonlinearity.
From this equation it is clear that
4-waves resonance conditions are satisfied if the 
frequency detuning $\Delta \omega_4$ of these 4-waves remains small:
\begin{equation}
\Delta \omega_4 = f(m+m_1-m'-m_1')+
\epsilon_m+\epsilon_{m_1} - \epsilon_{m'}-\epsilon_{m_1'} < \delta \omega 
\label{eq3}
\end{equation}
Thus the spreading over the ladder can proceed only over such modes where
$m+m_1-m'-m_1'=0$ and thus
$\Delta \omega_4= \epsilon_m+\epsilon_{m_1} - \epsilon_{m'}-\epsilon_{m_1'}$.
Since all $\epsilon_m$ are inside the frequency band $B$
it is possible to have $\Delta \omega_4 \sim 1/\Delta n$
so that the resonant detunings will remain small compared to
nonlinear shift $\delta \omega \sim \beta/\Delta n$ even at 
large $\Delta n$ values. 

The rate of spreading is still determined
by the same estimates as in \cite{dls1993,dls2008,dls2009} since
we still have $d C/dt \sim \beta C^3$ and the theoretical exponent is 
$\alpha = 2/5$ being independent of $f$ if $f \ell < 1$
so that the local transition rates remain the same as at $f=0$.
The numerical results presented here show weak dependence 
of $\alpha$ on $f$ for $f<1/\ell$ being in a satisfactory agreement
with this theoretical estimate. The deviation of $\alpha \approx 0.3$
from the theoretical value $2/5$ still should be better clarified
both for $f=0$ and $|f| > 0$.

In summary, we demonstrated that, in presence of a static field
applied to a lattice with disorder, a  nonlinearity
still produces complete delocalization with
a subdiffusive spreading over the ladder.
The exponent of the spreading remains close to the value 
without the force. The spreading forms
a homogeneous distribution of probability
inside a certain plateau.
Due to the conservation of total probability
and energy the far away parts of this
plateau remain statistically entangled even being  a large 
distance  apart from each other. The obtained results can be tested 
in experiments with nonlinear photonic lattices and BEC 
atoms in optical lattices with a static field.

\acknowledgments
We thank A.S.Pikovsky for useful discussions.
This research is supported in part 
by the ANR PNANO France project NANOTERRA.

\end{document}